\documentclass[epsfig,12pt]{article}
\usepackage{epsfig}
\usepackage{graphicx}
\usepackage{array}
\usepackage{color}
\usepackage{bm}
\usepackage{cite}
\usepackage{geometry}
\usepackage{latexsym}
 \usepackage{amsmath, amssymb, amscd, xypic, graphicx}

\newcommand{\beq}{\begin{equation}}   
\newcommand{\eeq}{\end{equation}}
\newcommand{\beqn}{\begin{eqnarray}}   
\newcommand{\eeqn}{\end{eqnarray}}

\newcommand*\xbar[1]{%
 \kern0.5ex%
  \hbox{%
   \kern0.2ex%
      \vbox{%
      \hrule height 0.5pt 
      \kern0.5ex
      \hbox{%
        \kern-0.1em
        \ensuremath{#1}%
        \kern-0.1em
      }%
    }%
  }%
}

\newcommand{\gsim}{\lower.7ex\hbox{$
\;\stackrel{\textstyle>}{\sim}\;$}}
\newcommand{\lsim}{\lower.7ex\hbox{$
\;\stackrel{\textstyle<}{\sim}\;$}}
\begin{document}

\begin{titlepage}

\begin{flushright}
FTPI-MINN-17/21, UMN-TH-3705/17
\end{flushright}

\vspace{5mm}

\begin{center}
{  \Large \bf  
Index Theorem for non-Supersymmetric Fermions\\[1mm]
 Coupled to a non-Abelian String and Electric\\[3mm]  Charge Quantization}
 
\vspace{5mm}

{\large \bf   M.~Shifman$^{\,a,b}$ and \bf A.~Yung$^{\,\,a,c,d}$}
\end {center}

\begin{center}

$^a${\it  William I. Fine Theoretical Physics Institute,
University of Minnesota,
Minneapolis, MN 55455}\\
$^{b}$ {\it Kavli Institute for Theoretical Physics, UC Santa Barbara} \\
$^{c}${\it National Research Center ``Kurchatov Institute,''
Petersburg Nuclear Physics Institute, Gatchina, St. Petersburg
188300, Russia}\\
$^{d}${\it  St. Petersburg State University,
 Universitetskaya nab., St. Petersburg 199034, Russia}
\end{center}

\vspace{1cm}


\begin{center}
{\large\bf Abstract}
\end{center}

Non-Abelian strings are considered in {\em non}-supersymmetric theories with fermions in various 
appropriate representations
of the gauge group U($N$). We derive the electric charge quantization conditions and the index theorems 
counting fermion zero modes in the string background both for the left-handed and right-handed fermions.
In both cases we observe a non-trivial $N$ dependence.

\vspace{2cm}

\end{titlepage}

\section{Introduction}

Non-Abelian (NA) vortex strings discovered in 2004 \cite{HT1,ABEKY,SYmon,HT2} in some supersymmetric Yang-Mills 
theories in four dimensions 
led to a large number of novel applications, see e.g.  \cite{Trev,Jrev,rev,rev22}. 
So far NA strings were explored mostly in the context of supersymmetric Yang-Mills theories. In this case the $U(1)$ charges of the relevant fermion matter fields 
(to be referred to as {\em electric charges}) coincide with those of the scalar fields whose winding insures topological stability of the NA vortices.  At the same time, NA strings are supported by some non-supersymmetric theories too \cite{GSY}.
In this case one can introduce a wider class of fermions which  in perturbation theory could have (in principle) arbitrary electric 
charges. We observe that the very existence of the NA vortex strings forces them to be quantized. 
This paper is devoted to two issues which were left behind in the studies of NA strings:
(i) electric charge quantization; (ii) 
the index theorem for the number of the fermion zero modes generalizing an analogous theorem \cite{JR,symindex} which is well-known from the 1970s. Due to a drastic difference between topology in the Abrikosov-Nielsen-Olesen  (ANO) and NA strings, respectively, both results to be presented below exhibit  quite remarkable $N$ dependence (here $N$ is the number of colors), see
Eqs. (\ref{sb15}), (\ref{D14}), and (\ref{D15}). 

The electric charge quantization is based on the Dirac argument. Dirac considered a $2\pi$ rotation of a probe electric charge around the Dirac string attached to the monopole and found that
\beq
{\cal M} \times Q = 2\pi
\eeq
were ${\cal M}$ is the magnetic flux associated with the Dirac monopole, and $Q$ is the electric charge. In what follows we will assume that
${\cal M}$ is measured in the units of $\frac 1e$ while $Q$ is measured in the units of $e$ defined in the Lagrangian through the coefficient of the
photon kinetic term.

The index theorems for NA strings in non-supersymmetric models we derived are presented in Eqs. (\ref{sb22}) and  (\ref{sb23}).

\section{Non-supersymmetric  model with NA strings}
\label{TWO}

The basic model (prior to introduction of fermions) has U($N$) gauge group with gauge potentials $A_{\mu}$ and  $A^a_{\mu}$,
 where $a$ is the adjoint index of the color $SU(N)$. The action has the form
\beq
{\cal L}_0 = -\frac{1}{4g^2} F_{\mu\nu}^a  F_{\mu\nu}^{a}  -\frac{1}{4e^2} F_{\mu\nu}  F_{\mu\nu} + \sum_A\left| {D_\mu} \phi^A \right|^2 - V(\phi)\,,
\label{fr1}
\eeq
where the scalar fields $\phi^A$ for each $A$ are  the fundamental representation of the color $SU(N)$, with suppressed color indices $i=1,2,... , N$. The subscript $A$ labels the scalar field flavors, $A= 1,2,... , N$ (i.e. in this case $N_c=N_f$). 
The covariant derivative is defined as follows
\beq
D_\mu
=\partial_\mu- \frac i2 A_\mu- i A_\mu^a \,T^a\,,\qquad i=1,2\,,
\label{10}
\eeq
where the matrices $T^a$ are the generators of the gauge $SU(N)$.

The scalar sector is assumed to have a global (flavor) $SU(N)$ symmetry so that the potential $V(\phi )$ must be 
chosen appropriately. 
Our choice will be motivated by supersymmetric field theory (see e.g. \cite{rev}), 
\beq
V= h_1 \sum_a\left(\sum_A \bar\phi_A T^a \phi^A\right)^2+h_2 \left(\sum_a \bar\phi_A  \phi^A - Nv^2\right)^2 \,.
\label{fr2}
\eeq
Here  $h_{1,2}$ are constants. For what follows it is convenient to introduce $N\times N$ matrix $\Phi = \{ \phi_i^A\}$ which is constructed of $N$ columns
\beq
\Phi \leftrightarrow \{ \phi_i^1,\, \phi_i^2, \, ... , \phi_i^N\}
\label{fr3}
\eeq
where the superscript marks flavor while the subscript $i$ is the color index in  the fundamental representation, 
$i=1,2,..., N$.
In this notation
\beqn
{\cal L}_0 &=& -\frac{1}{4g^2} F_{\mu\nu}^a  F_{\mu\nu}^{a} -\frac{1}{4e^2} F_{\mu\nu}  F_{\mu\nu}
 + {\rm Tr}\, \left( {D_\mu} \Phi \right)^\dagger
\left( {D_\mu} \Phi \right) -V(\phi)\,,\nonumber\\[3mm]
V&=& \lambda_1 \sum_a\left( {\rm Tr}\, \Phi^\dagger T^a \Phi\right)^2
+\lambda_2 \left[ {\rm Tr }\left( \Phi^\dagger  \Phi - Nv^2\right) \right]^2 \,.
\label{fr4}
\eeqn

Now, it is perfectly clear that  $\Phi$ develops the diagonal expectation value of the form
\beq
\Phi_{\rm vac} =\left(\begin{array}{cccc}
v&0&...&0\\[1mm]
0&v&...&0\\[1mm]
...&...&...& ...\\[1mm]
0&0&...&v
\label{fr5}
\end{array}
\right)
\eeq
 (up to irrelevant gauge transformations). The parameter $v$ can always be chosen real because of the $U(1)$ gauge symmetry obvious in (\ref{fr1}).

 This set-up (first considered in the context of supersymmetric vortices in ${\cal N}=2$ Yang-Mills theories)
leads to non-Abelian flux tubes in non-supersymmetric theories  \cite{GSY}. Non-Abelian flux tubes (strings) are  described in detail in \cite{rev}.
  
 All $N^2$ gauge bosons of the original $U(N)$ gauge theory are Higgsed, acquiring  masses $M=gv$, $M=ev$, while $N^2$
 phases are eaten up by gauge bosons and  all $N^2$ real physical Higgses acquire masses.

Equation (\ref{fr5}) shows that in the Higgs regime both the local gauge and global flavor groups are spontaneously
broken, but the diagonal $SU(N)$ survives as the {\em exact global} symmetry of the model. This is called color-flavor locking.

\section{Adding fermions}

One can consider two distinct types of fermions: those which  have 
no couplings to the field $\Phi$ and, the second class, 
those coupled to $\Phi$ via Yukawa couplings. For simplicity we will assume that the fermions from the first class have only electric charges (i.e. not coupled to the non-Abelian field) and call them {\em leptons}. The fermions from the second class will be referred as {\em quarks}. Both classes can be expanded in a straightforwards manner.

Before we address the issue of the electric charge quantization, let us say a few words about the structure of the non-Abelian  strings.
We need to know the flux of the Abelian magnetic field flowing through the string, as it differs from that
in the ANO string.

How is topological stability implemented for the non-Abelian string? The ANO strings are stable because $\pi_1 (U(1))=Z$. This is not the case in the 
non-Abelian string.
The key point is that the center elements of the $SU(N)$ group are simultaneously elements of $U(1)$.

Imagine a large circle in the plane perpendicular to the string axis with the center coinciding with the center of the string core. In the case of the Abelian string, when we travel along this circle making the full rotation,  we travel from $1$ to $e^{2\pi i }$  in the $U(1)$ group space. In the case of non-Abelian string with, say,  the $U(2)$ gauge group 
we travel in $U(1)$ from $1$ to $e^{\pi i}$	and then the remaining necessary interpolation is realized through the $SU(2)$ factor group (in Eq. (\ref{15}) we apply the $\sigma_3$ rotation),
\beq
\phi \to  \underbrace{ \left(\begin{array}{c} e^{\pi i}\\[2mm] e^{\pi i}\end{array}
\right)}_{\text{$U(1)$}} \times \underbrace{ \left(\begin{array}{c} e^{\pi i}\\[2mm] e^{-\pi i}\end{array}
\right)}_{\text{$SU(2)$}} \phi\,.
\label{15}
\eeq

In the minimal non-Abelian string in the $U(N)$ theory the U(1) interpolation connects 
two nearest center elements of $SU(N)$, namely, say, unity and $\exp(\frac{2\pi i}{ N})$.
Hence, the  Abelian field magnetic flux for the minimal non-Abelian string is
\beq
\int \,d^2 x \left( \frac 12 \varepsilon_{ij} F_{ij}\right) = \frac{4\pi}{N}\,.
\label{D8}
\eeq
For non-minimal strings we can have
\beq
\int \,d^2 x \left( \frac 12 \varepsilon_{ij} F_{ij}\right) = \frac{4\pi k}{N}\,,\qquad k=1,2,...,\,N-1.
\label{D9}
\eeq
Note that the numerators on the right-hand sides in (\ref{D8}) and (\ref{D9}) contain the factors $4\pi$, rather than the conventional $2\pi$. 
This is due to the fact that we normalize the covariant derivative acting on $\phi$ as in Eq. (\ref{10}), i.e. the electric charge of the $\phi$ field is $Q(\phi)=
\frac 12$, rather than $1$. 

We add leptons $\ell_\alpha$ with the electric charge $Q(\ell)$ to be determined below. The leptons are assumed to be Dirac, so adding mass terms is possible.

The quarks must be chosen in such a way -- and coupled to the Higgs fields -- so that they all become massive upon Higgs condensation.
This will require considering two color indices for some fermions. Note that three-index fermion representations are not appropriate at arbitrary $N$ since they will ruin asymptotic freedom. With this remark in mind the choice is unique:
\beq
\psi^{ij}_\alpha\,,\,\, \tilde{\psi}_{ij}^\alpha \,,\,\, \left(\eta_\alpha\right)_{i}^A 
\,,\,\,\left({\tilde\eta}^\alpha\right)^{i}_A \,,
\label{sixpp}
\eeq
where $\alpha$ is the Weyl Lorentz index. Note that two-index representations $\psi^{ij}$ and $\tilde{\psi}_{ij}$ are reducible since we do not symmetrize (anti-symmetrize) the indices. Separating symmetric from antisymmetric representation is possible but will unnecessary complicate the argument by a technicality.

All Weyl spinor fields in (\ref{sixpp})
are left-handed; their complex conjugated (right-handed) fields with dotted spinor indices are also introduced.

Then the Yukawa interaction takes the form
\beq
V_{\rm Yukawa} = \left(\lambda\, \bar{\phi}_{Ai}\eta_{j}^A\psi^{ji}   + \tilde\lambda\, \tilde\psi_{ij} \tilde\eta^{j}_A
 \phi^{iA}\right) + {\mbox{H.c.}}
\label{sevenp}
\eeq
The Lorentz indices in (\ref{sevenp}) are contracted as $\psi\eta \equiv \varepsilon^{\alpha\beta} \psi_\alpha\eta_\beta$ and so on, according to the standard four-dimensional rule. Obviously,  the expectation value (\ref{fr5}) generates mass terms to all quark fields introduced above. 
One can additionally introduce massless quark fields decoupled from $\phi$. These latter quarks can be massive provided they are represented by the Dirac fields.

We summarize representations with respect to color and flavor group for fields $\phi$, $\psi$ and $\eta$ in the 
Table \ref{table1}.

\begin{table}
\begin{center}
\begin{tabular}{|c|c | c| c|c| c |}
\hline
$\rule{0mm}{6mm}$ Field & $\phi$ & $\psi$ & $\tilde{\psi}$ & $\eta$ & $\tilde{\eta}$
\\[3mm]
\hline
$\rule{0mm}{5mm}$ SU$(N)_G$  & $N$ &  $N^2$ & $\bar{N}^2$ & $\bar{N}$ & $N$ 
\\[2mm]
\hline
$\rule{0mm}{5mm}$ SU$(N)_{F}$  & $N$ & 1 & 1 & $N$ & $\bar{N}$
\\[2mm]
\hline
$\rule{0mm}{5mm}$ U(1)$_G$  & $\frac12$ & $Q_{\psi}$ & $-Q_{\psi}$ & $Q_{\eta}$ & $-Q_{\eta}$
\\[2mm]
\hline
\end{tabular}
\end{center}
\caption{{\small  Gauge and global representations of fields $\phi$, $\psi$ and $\eta$. The last line
presents their electric charges. }}
\label{table1}
\end{table}

\section{Electric charges}

We define electric charges  $ Q(\psi) $ and  $Q(\eta )$ by the U(1) parts of the respective covariant derivatives,
\beq
{\cal D}_\mu \eta
=\big[\partial_\mu-  i\,Q(\eta ) A_\mu + ... \big] \eta \,,\qquad {\cal D}_\mu \psi
=\big[\partial_\mu-  i\,Q(\psi ) A_\mu+...\big] \psi
\eeq
The normalizations of the charge $Q( \phi )$
is determined by the Abelian part of the covariant derivative acting on $\phi$, see \eqref{10}.
 With this convention  
\beq
Q(\phi) = \frac 12\,,
\label{11}
\eeq
as was stated above. Charges of the fermion fields in our model are shown in Table \ref{table1}.

Now let us determine possible electric charges of leptons. Consider the minimal string (\ref{D8}), draw a large circle in the plane perpendicular to the NA string axis,
and drag your probe particle along this contour. The Aharonov-Bohm phase of the corresponding wave function should be
 $\pm 2\pi$, $\pm 4\pi$, etc.

This implies that the minimal lepton electric charge satisfying the condition ${\cal M} Q =2\pi $ is as follows:
\beq
| Q(\ell)| = \frac{N}{2}\,.
\label{sb15}
\eeq
In particular, if the number of colors $N$ is large, so is the lepton electric charge.

Now, let us pass to the quarks. The simplest way to derive the electric charge quantization is to consider
 color-singlet ``hadrons," for instance, $\eta_{i}^{A} \phi^{iB}$ or 
$ \psi^{ij} \bar{\phi}_{Ai}\bar{\phi}_{Bj}$. If we drag these hadrons along the large circle, as above, we will see that their electric charges can be $0$, $\pm \frac{N}{2}$, etc., just as for the leptons. Considering the former case we conclude that
\beq
Q(\eta ) = -\frac 12\,,\,\, \mp \frac N2 - \frac 12\,,\,\, \mp N  - \frac 12 \,, ...\,.
\label{D14}
\eeq

Then, for $Q(\psi )$ we obtain
\beq
Q(\psi ) = 1\,,\,\, \pm \frac N2 +1\,,\,\, \pm N  +1 \,, ...\,.
\label{D15}
\eeq

The correlation of the plus-minus signs in (\ref{D14}) and (\ref{D15}) is due to the electric charge conservation of the Yukawa term (\ref{sevenp}),
\beq
Q(\psi) + Q(\eta ) =  Q( \phi )=\frac12\,,\qquad Q(\tilde\psi) + Q(\tilde\eta ) =  Q( \bar\phi )=-\frac12\,.
\label{eight}
\eeq

Aharonov-Bohm phases for NA strings in theories with gauged color-flavor group were discussed  in \cite{Konishi15}.
These phases were studied for NA semilocal strings in SQCD in \cite{studKomarg}.

\section{Index theorem}

Here we will find the index of the two-dimensional Dirac operator in the background of the NA string. The Dirac operator becomes two-dimensional because
the string solution depends only on $x$ and $y$ and is $t,z$ independent.

We must reduce 4D fermions into 2D fermions.
In two dimensions, the distinction between the dotted and undotted indices becomes irrelevant;  they do not correspond to 2D chirality.
In fact, all four fields in (\ref{sixpp}) (combined with their complex conjugated) become Dirac fermions. The two-dimensional reduction of the theory
has four (rather than two) Dirac fermions. Hence,
we can identify  four independent 2D axial currents:
\beq
j_\mu^5 = \left\{\bar\psi \gamma_\mu\gamma_5\psi\,,\quad \bar{\tilde\psi} \gamma_\mu\gamma_5\tilde\psi\,\quad
\bar\eta \gamma_\mu\gamma_5\eta\,,\quad \bar{\tilde\eta} \gamma_\mu\gamma_5\tilde\eta
\right\},\qquad \mu = 1,2\,,
\eeq
where $\gamma_{1,2} = \sigma_{1,2}$ and $\gamma_5=\sigma_3$.

Taking into account Yukawa interactions in Eq. (\ref{sevenp}) we see that
only two of axial symmetries are classically conserved. Let us call them U(1)$_L$ and U(1)$_R$. Fields $\psi$ and
$\eta$ have the same charges with respect to U(1)$_L$ while $\tilde{\psi}$ and $\tilde{\eta}$ are neutral. Vice verse
fields $\tilde{\psi}$ and $\tilde{\eta}$ have the same charges with respect to U(1)$_R$, while $\psi$ and
$\eta$ are neutral. Thus we have two classically conserved currents
\beqn
J_\mu^{5} = j_\mu^5(\psi) +  j_\mu^5(\eta)\,,\qquad {\tilde J}_\mu^5 = j_\mu^5(\tilde\psi) +  j_\mu^5(\tilde\eta)\,
\eeqn
associated with U(1)$_L$ and U(1)$_R$.

However, they are 2D anomalous with regard to the photon (but {\em not} with regard to non-Abelian gauge bosons).
In two dimensions the anomalous loop is {\em diangle}, rather than triangle, and, therefore, is proportional to $Q$. 
Namely,
\beqn
\partial_i J_i^5 = \big[Q(\psi) +Q(\eta)\big]\,  \frac{N^2}{2\pi} \varepsilon_{ij}F_{ij}\,, \nonumber\\[2mm]
\partial_i  {\tilde J}_i^5 =\big[Q(\tilde\psi) +Q(\tilde \eta)\big]\,  \frac{N^2}{2\pi} \varepsilon_{ij}F_{ij}\,.
\label{14}
\eeqn

\vspace{3mm}

Integrating both sides over $d^2x$ and combining Eqs. (\ref{D8}) and (\ref{eight}) we 
see that  the NA string induces a change in U(1)$_L$ charge, namely, 
\beq
\delta Q_{L} = 4N\big[Q(\psi) +Q(\eta)\big]=2N
\label{sb22}
\eeq
while the change in U(1)$_R$ charge is  as follows:          
\beq
\delta Q_{R} = -4N\big[Q(\psi) +Q(\eta)\big]=-2N \, .
\label{sb23}
\eeq
Thus each fermion system  $\psi$-$\eta$ and $\tilde{\psi}$-$\tilde{\eta}$ has $2N$ zero modes in the background of 
the NA string.

The index theorem for NA string in supersymmetric setting was considered in \cite{rev},
see also \cite{Nitta}.

\section*{Acknowledgments}

This work  is supported in part by DOE grant DE-SC0011842. 
The work of M.S.  was supported in part by the National Science Foundation under Grant No. NSF PHY-1125915. 
M.S. is grateful to KITP where this paper was completed for kind hospitality extended to him during the workshop ``Resurgence 17."
The work of A.Y. was  supported by William I. Fine Theoretical Physics Institute  at the  University 
of Minnesota, by the Russian Foundation for Basic Research grant
No.~18-02-00048 and by Russian State Grant for
Scientific Schools RSGSS-657512010.2.


\begin{thebibliography}{99}

{\small


\bibitem{HT1}
A.~Hanany and D.~Tong,
``Vortices, instantons and branes,''
JHEP {\bf 0307}, 037 (2003)
[hep-th/0306150].

\bibitem{ABEKY}
R.~Auzzi, S.~Bolognesi, J.~Evslin, K.~Konishi and A.~Yung,
{\em Non-Abelian superconductors: Vortices and
confinement in N = 2 SQCD,}
Nucl.\ Phys.\ B {\bf 673}, 187 (2003)
[hep-th/0307287].

\bibitem{SYmon}
M.~Shifman and A.~Yung,
``Non-Abelian string junctions as confined monopoles,''
Phys.\ Rev.\ D {\bf 70}, 045004 (2004)
[hep-th/0403149].

\bibitem{HT2}
A.~Hanany and D.~Tong,
``Vortex strings and four-dimensional gauge dynamics,''
JHEP {\bf 0404}, 066 (2004)
[hep-th/0403158].


\bibitem{Trev}
D.~Tong,
{\em TASI Lectures on Solitons,}
  arXiv:hep-th/0509216.

\bibitem{Jrev}
  M.~Eto, Y.~Isozumi, M.~Nitta, K.~Ohashi and N.~Sakai,
  J.\ Phys.\ A  {\bf 39}, R315 (2006)
  [arXiv:hep-th/0602170];
K.~Konishi,
  Lect.\ Notes Phys.\  {\bf 737}, 471 (2008)
  [arXiv:hep-th/0702102];

    \bibitem{rev}
  M. Shifman, A. Yung, {\sl Supersymmetric Solitons}, 
  (Cambridge University Press, 2009) [ Rev.\ Mod.\ Phys.\  {\bf 79}, 1139 (2007)
  [hep-th/0703267]]; {\em Critical String from Non-Abelian Vortex in Four Dimensions,}
  Phys.\ Lett.\ B {\bf 750}, 416 (2015)
  [arXiv:1502.00683 [hep-th]].
  
      \bibitem{rev22}
       M.~Shifman and A.~Yung,
{\em Lessons from supersymmetry: ``Instead-of-Confinement" Mechanism,}
  Int.\ J.\ Mod.\ Phys.\ A {\bf 29}, no. 27, 1430064 (2014)
  [arXiv:1410.2900 [hep-th]].

\bibitem{GSY}
A.~Gorsky, M.~Shifman and A.~Yung,
 {\em Non-Abelian Meissner effect in Yang-Mills theories at weak coupling,}
  Phys.\ Rev.\ D {\bf 71}, 045010 (2005)
  [hep-th/0412082].
  
\bibitem{JR}
  R.~Jackiw and P.~Rossi,
 {\em Zero Modes of the Vortex-Fermion System,}
  Nucl.\ Phys.\ B {\bf 190}, 681 (1981).
  
  \bibitem{symindex}
M.  Shifman, {\sl Advanced Topics in Quantum Field Theory}, (Cambridge University Press, 2012).}
  
  \bibitem{Konishi15} 
{Stefano~Bolognesi, C.~Chatterjee, and Ken~Konishi,} 
{\em Non-Abelian Vortices, Large Winding Limits and Aharonov-Bohm Effects,} 
JHEP \ {\bf 1504}, 143 (2015) 
[arXiv:1503.00517].

\bibitem{studKomarg}
E.~Gerchkovitz and  A.~Karasik, 
{\em Vortex-Strings in N=2 SQCD and Bulk-String Decoupling,}
 [arXiv:1710.02203].
 
   \bibitem{Nitta}
   T.~Fujiwara, T.~Fukui, M.~Nitta and S.~Yasui,
{\em Index theorem and Majorana zero modes along a non-Abelian vortex in a color superconductor,}
  Phys.\ Rev.\ D {\bf 84}, 076002 (2011)
  [arXiv:1105.2115 [hep-ph]].



\end{thebibliography}
\end{document}